\documentclass[conference]{IEEEtran}

\IEEEoverridecommandlockouts

\usepackage[absolute]{textpos}
\newcommand{\copyrightstatement}{
    \begin{textblock}{15}(0.5,0.3)    % tweak here: {box width}(leftposition, rightposition)
         \noindent
         \centering
         \textblockcolour{white}
         \footnotesize
         \copyright 2019 IEEE. Personal use of this material is permitted. Permission from IEEE must be obtained for all other uses, in any current or future media, including reprinting/republishing this material for advertising or promotional purposes, creating new collective works, for resale or redistribution to servers or lists, or reuse of any copyrighted component of this work in other works
    \end{textblock}
}

\usepackage{cite}

  \usepackage{graphicx}                % allow us to embed graphics files
  \DeclareGraphicsExtensions{.eps}  
  
\usepackage{amsmath}
\usepackage{psfrag}

\usepackage{array}

\usepackage{url}

\begin{document}

\title{ Power-Efficient Video Streaming on Mobile Devices Using Optimal Spatial Scaling}

\copyrightstatement

\author{
\IEEEauthorblockN{Christian Herglotz, Andr\'e Kaup (\textit{Fellow IEEE})}
\IEEEauthorblockA{Multimedia Communications \& \\ Signal Processing, \\
Friedrich-Alexander University \\ Erlangen-N\"urnberg (FAU)\\
Erlangen, Germany}

\and 

\IEEEauthorblockN{St\'ephane Coulombe}
\IEEEauthorblockA{ Dept. of Software Engineering and IT\\
\'Ecole de technologie sup\'erieure (\'ETS)\\
Montr\'eal, Canada}

\and

\IEEEauthorblockN{Ahmad Vakili}
\IEEEauthorblockA{Summit Tech Multimedia\\
Montr\'eal, Canada\\
www.summit-tech.ca}

\vspace{-0.5cm}
%\thanks{
%\noindent\hrulefill
%\par
% }%thanks

}%author

% For invited papers. 

\maketitle

\begin{abstract}
This paper derives optimal spatial scaling and rate control parameters for power-efficient wireless video streaming on portable devices. A video streaming application is studied, which receives a high-resolution and high-quality video stream from a remote server and displays the content to the end-user. We show that the resolution of the input video can be adjusted such that the quality-power trade-off is optimized. Making use of a power model from the literature and subjective quality evaluation using a perceptual metric, we derive optimal combinations of the scaling factor and the rate-control parameter for encoding. For HD sequences, up to $10\%$ of power can be saved at negligible quality losses and up to $15\%$ of power can be saved at tolerable distortions. To show general validity, the method was tested for Wi-Fi and a mobile network as well as for two different smartphones. 

\end{abstract}

\begin{IEEEkeywords}
video coding, resolution, low-power, smartphone, video streaming, VMAF
\end{IEEEkeywords}

% \fi
\IEEEpeerreviewmaketitle

\section{Introduction}
During the past years, video streaming applications have conquered the mass markets such that nowadays, watching videos online can be performed with many portable devices, such as a smartphone or a tablet PC. Wireless networks using standards like Wi-Fi, 4G, or the upcoming 5G allow the streaming of high-quality video content at high resolutions. However, a major problem in video streaming is the power consumption of the portable device, because the streaming application can drain the battery quickly. As a consequence, operating times during streaming tend to be short and the consumer's quality of experience (QoE) is strongly impaired. 

To tackle this problem, we take a close look at the power consumption of typical smartphones during video streaming. From the literature, we know that the bitrate and the resolution have a major influence on the power consumption in such a way that a lower bitrate and a lower resolution decrease the streaming power significantly \cite{Herglotz18a}. On the other hand, decreasing both values also impairs the visual quality of the displayed video. Therefore, in this work, we perform a thorough analysis of the power consumption and the visual quality at different resolutions as well as bitrates and propose parameter combinations for optimal power-quality trade-offs. It is worth mentioning that in this context, we assume that the bitrate of the video stream is significantly smaller than the capacity of the transmission channel such that the bitrate does not have to be considered. 

\begin{figure}[t]
\centering
\psfrag{P}[c][c]{Power}
\psfrag{M}[c][c]{Analysis}
\psfrag{D}[c][c]{Portable}
\psfrag{S}[c][c]{Device}
\psfrag{R}[c][c]{Streaming Server}
\includegraphics[width=.4\textwidth]{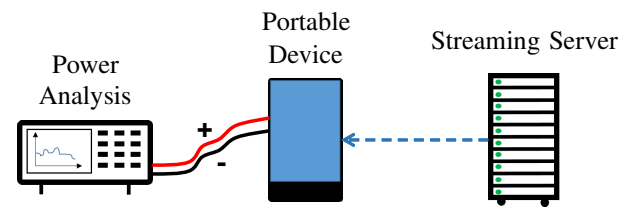}
\vspace{-.2cm}
\caption{Power consumption analysis for a portable device during wireless video streaming. The stream is provided by a remote streaming server. }
\label{fig:setup}
\end{figure}

To this end, we consider the power consumption of a smartphone that runs a video streaming application (cf. Fig.\ \ref{fig:setup}). The streaming video is coded with a common compression format like H.264/AVC \cite{ITU_H.264} or HEVC \cite{ITU_HEVC} and streaming is performed via a wireless network. In this work, we restrict our considerations to H.264/AVC because, in 2018, this codec still accounted for more than three-thirds of all coded video data worldwide \cite{statistaH264}.  % or the stream read from the local memory of the device. 
The power is considered to be the overall power of the device, which can be measured using the battery connectors. As a consequence, we consider the overall power consumption including network receiver power, decoding, rendering, and display. The application receives the stream, decodes the stream using an internal hardware decoder, and performs upsampling to obtain a full-screen output. 

For the evaluation of the visual quality of the output video, we use an objective and a perceptual quality metric, which is trained on subjective data. 
%the common objective metric Peak Signal-to-Noise Ratio (PSNR), which is related to the mean square error (MSE). Furthermore, we employ video multimethod assessment fusion (VMAF) \cite{VMAF}, which was designed to reflect the quality experience of the viewer using machine learning methods. 
For exhaustive power considerations, next to true power measurements, we exploit a video decoding energy model from the literature \cite{Herglotz18a}, which was successfully employed to model virtual reality streaming applications \cite{Herglotz19a}. 

The method of reducing the resolution for optimizing the compression efficiency in video coding has been studied thoroughly in the literature. Wang et al. \cite{Wang14} performed an analysis of the distortion induced by downsampling and found that it is barely correlated to coding artifacts caused by quantization. As a consequence, they proposed a method to adaptively downsample image content to increase the compression efficiency at low visual qualities. In a similar direction, Afonso et al. \cite{Afonso17} proposed an adaptive frame-based downsampling scheme and obtained bitrate reductions of approximately $4\%$ at the same objective visual quality. Further work showed that using neural networks for image enhancement, one can increase the quality of images distorted by downsampling \cite{Jenab18, Hamis19}. For practical applications, Dragi\'c et al. showed that power can be saved by adapting the video resolution to the device's display \cite{Dragic14}. 

In terms of the power consumption, Li et al. constructed a power model for H.264/AVC streaming and proposed to optimize the power-rate costs \cite{Li12}. However, resolution downscaling was not considered. In a similar direction, a method for optimizing the software decoding energy using decoding-energy-rate-distortion optimization was proposed in \cite{Herglotz19}. In contrast, in this work we tackle hardware decoding and propose to exploit both resolution downsampling and quantization control for optimal visual qualities at a high power efficiency. %, independent from the resolution of the output display. 

This paper is organized as follows. First, Section \ref{sec:simul} summarizes the simulation setup used for power and quality assessment. Afterwards, Section \ref{sec:eval} discusses the simulation results in detail and provides recommendations for optimal coding decisions. Finally, Section \ref{sec:concl} concludes this paper. 

\section{Simulation Framework}
\label{sec:simul}
\begin{figure}[t]
\centering
\psfrag{P}[c][c]{Power}
\psfrag{M}[c][c]{Modeling}
\psfrag{D}[c][c]{Downsampling}
\psfrag{S}[c][c]{Measurement}
\psfrag{R}[c][c]{$S$}
\psfrag{b}[c][c]{$b$}
\psfrag{s}[c][c]{$s$}
\psfrag{C}[c][c]{$c$}
\psfrag{E}[c][c]{Encoder}
\psfrag{H}[c][c]{Decoder}
\psfrag{U}[c][c]{Upsampling}
\psfrag{A}[c][c]{Evaluation}
\psfrag{Q}[c][c]{Quality}
\includegraphics[width=.49\textwidth]{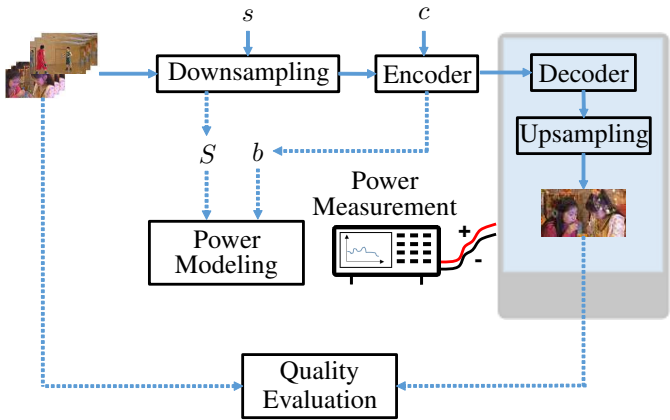}
\vspace{-.2cm}
\caption{Framework for the assessment of the mobile device's power consumption and the visual quality of the output video.  }
\label{fig:framework}
\end{figure}

A block diagram of our evaluation framework is depicted in Fig.\ \ref{fig:framework}. 
At first, we downsample the input sequence using bicubic interpolation. In order to do so, we use FFmpeg's internal spatial scaling algorithm, in which the target pixel width $w_\mathrm{target}$ and pixel height $h_\mathrm{target}$ can be freely chosen \cite{FFmpeg}. To determine the target resolution, we use the downscaling factor $s$ which can be used to calculate the target resolution by 
\begin{align}
w_\mathrm{target} = & \mathrm{round}\left(\frac{w_\mathrm{in}\cdot s}{4}\right)\cdot 4, \\
h_\mathrm{target} = & \mathrm{round}\left(\frac{h_\mathrm{in}\cdot s}{4}\right)\cdot 4, \label{eq:targetRes}
\end{align}
where $w_\mathrm{in}$ and $h_\mathrm{in}$ are the width and the height of the original input sequence, respectively. The rounding operation is performed to obtain an integer number of pixels. The division by four, as well as the multiplication by four, ensures that the output resolution is covered by the video codec standards, which only allow multiples of four as width and height. As a consequence, $s^2$ represents the fraction of pixels in relation to the number of pixels in the original sequence. For example, if $s^2=0.5$, the total number of pixels after downscaling is half as high as in the original resolution. 

Afterwards, the output sequence from downsampling is encoded with the standard encoder included in FFmpeg, i.e., x264 \cite{x264}. % and x264 for H.264 \cite{x264}. 
For the sake of simplicity, we do not change presets and only use the constant rate factor (crf) $c$ for bitrate control, which is generally proportional to the quantization parameter (QP). In contrast to the QP, the crf is supposed to keep the perceptual quality (not the objective quality) of the compressed sequence at a constant level. This is done by encoding frames including a lot of motion with a higher QP and frames with small motion at a smaller QP. Furthermore, using the output resolution $S = w_\mathrm{target}\cdot h_\mathrm{target}$ and the output bitrate $b$ from encoding, we are able to obtain power estimates from the power model that will be introduced in Subsection \ref{secsec:powerModel}. 

In addition to power modeling for exhaustive simulations, we performed several power measurements to support our findings (right of Fig.\ \ref{fig:framework}). To this end, we measure the power consumption of a smartphone using a power meter attached to its battery connectors. A video streaming application is executed which receives the video stream from a remote server or reads it from the local memory, decodes the stream using an on-chip hardware decoder, upsamples the video to the display's resolution $S_\mathrm{display}$, and shows the video on the screen. The measurement setup is the same as in \cite{Herglotz19a}. 

For quality evaluation of the output video (bottom of Fig.\ \ref{fig:framework}), we simulate decoding and upsampling in the portable device using FFmpeg's decoder and the same spatial scaling algorithm as used for downsampling. For the interpolation filter, we again choose bicubic interpolation because it can easily be implemented using on-chip graphics processing units (GPUs) and it has low complexity demands such that the corresponding power consumption on a portable device would be small. We upsample the decoded video to HD resolution and compare with the original sequence that was also upsampled to HD, where we keep the aspect ratio constant (that means, either the width or the height corresponds to HD). The quality metrics used in this work are explained in Subsection \ref{secsec:qual}.

\subsection{Power Modeling}
\label{secsec:powerModel}
To measure the impact of the resolution and other factors on the energy or power consumption during video streaming and coding, detailed investigations were performed and published in \cite{Herglotz18, Herglotz18a, Herglotz19a}. In particular, it was found that the power can be accurately modeled using a few parameters, which describe high-level properties of the bit stream. In this work, we take the power model for video decoding from \cite{Herglotz19a} that reads 
\begin{equation}
\hat P = p_\mathrm{0} + p_{f}\cdot f + p_{S}\cdot S + p_{b}\cdot b, 
\label{eq:PM}
\end{equation}
where the variable $f$ is the frame rate, $S$ is the resolution in pixels per frame, and  $b$ the bitrate of the video stream. The parameters $p_{f}$, $p_{S}$, and $p_{b}$ are model parameters that describe a linear relationship between the variables and the overall power, and $p_\mathrm{0}$ is a constant offset. Results reported in \cite{Herglotz19a} and \cite{Herglotz18a}, in which the model was first proposed for energy estimations, indicate that this model is capable of estimating the overall power with a mean error of less than $5\%$. 

In this paper, we exploit the fact that by changing the pixel dimensions of the frames and the quantization parameter during encoding, we can directly influence the overall power with the help of the two variables resolution $S$ and bitrate $b$ (cf. Fig.\ \ref{fig:framework}). Since the corresponding parameters $p_{S}$ and $p_{b}$ have positive values, we can decrease the overall power consumption. For resolution and bitrate control, we use the downsampling factor $s$ and the crf $c$. 

\subsection{Quality Evaluation}
\label{secsec:qual}
In order to evaluate the visual quality of the output sequence, we exploit two different metrics. The first metric is the PSNR which is an objective metric used to assess the video quality of compressed images. The PSNR is a full reference metric calculated by 
\begin{equation}
\mathrm{PSNR} = 10\cdot \mathrm{log}_{10}\left( \frac{M^2}{\mathrm{MSE}}\right), 
\label{eq:PSNR}
\end{equation}
where $M$ is the peak value a pixel can obtain ($M=255$ for $8$-bit sequences) and MSE the mean-square-error between the original sequence and the reconstructed sequence. The PSNR is calculated for each color component Y, U, and V and then averaged to the YUV-PSNR
\begin{equation}
\mathrm{PSNR}_\mathrm{YUV} = \frac{1}{8}\left(6\cdot \mathrm{PSNR}_\mathrm{Y} + \mathrm{PSNR}_\mathrm{U} + \mathrm{PSNR}_\mathrm{V}\right), 
\label{eq:YUVPSNR}
\end{equation} 
in which the luma component Y has the highest weight \cite{Ohm12}. The overall PSNR of a video is then obtained by calculating the mean YUV-PSNR over all frames. 

A major drawback of the PSNR is that it does not always reflect the perceptual quality of a video. To tackle this issue, a lot of research was performed to obtain metrics reproducing the perceptual quality \cite{Eskicioglu95,Wang04, Zhang17}. In this work, we consider the video multi-method assessment fusion (VMAF) \cite{VMAF} metric, which is a machine learning-based approach. The concept of VMAF is to use a support vector machine (SVM) that exploits the information from different quality metrics and maps them to subjective quality scores. The SVM is trained using a large database  from true subjective video quality assessments and it can be shown that VMAF provides high correlations with subjective scores. The subjective scores were obtained using the popular mean-opinion-score (MOS) and the VMAF score maps them to the range $[0, 100]$ with $100$ being the best score. For the target of mobile streaming in this paper, we use the VMAF-model version 0.6.1, which includes data that is explicitly trained for smartphones. It is available using the command line flag \texttt{--phone-model} \cite{VMAF}. 

\section{Evaluation}
\label{sec:eval}
The evaluation is split into three parts. In the first part, the general setup for power modeling is presented. In the second part, the results from the quality-power assessment are discussed in detail. In the third part, recommendations for practical applications are derived. 

\subsection{Power Modeling Setup}
\label{secsec:setup}
For the evaluation of the quality-power performance, we encode 20 different sequences from the HEVC common test conditions \cite{Bossen13}. The sequences are given in the resolutions $416\times 240$, $832\times 480$, $1280\times 720$, and $1920\times 1080$. The latter high-definition (HD) resolution is the largest resolution we consider because it is a common resolution of modern smartphone displays. The sequences mainly contain natural content including camera movement (sports, public places, indoor), three sequences were recorded with a static camera similar to videoconferencing, and three sequences show screen content. 

The power evaluation is performed for two modern smartphones with the properties listed in Table\ \ref{tab:specs}. %The first includes a quadcore system-on-chip with hardware video decoding and a liquid crystal display (LCD) of HD resolution. The main specifications are listed in Table\ \ref{tab:specs}. 
For online streaming, a Wi-Fi connection was established and the video was streamed using a real-time messaging protocol (RTMP). In addition, 3G streaming was tested for device A. Hence, three different cases are evaluated, namely A-Wi-Fi, A-3G, and B-Wi-Fi.  
%\begin{table}[t]
%\renewcommand{\arraystretch}{1.3}
%\caption{Main specifications of the tested DUT. The display uses the in-plane switching (IPS) technology for thin-film-transistor (TFT) LCDs.  }
% \vspace{-0.4cm}
%\label{tab:specs}
%\begin{center}
%\begin{tabular}{l|l}
%\hline
%\begin{tabular}{l}
%CPU \\
%GPU\\
%Memory\\
%Multimedia\\
%Display\\
%\end{tabular}
%& \hspace{-0.35cm}
%\begin{tabular}{l}
% Four $32$ bit cores (max $2.5$ GHz)\\
% Up to $578$ MHz, $147.9$ GFLOPS\\
%$2$ GB of LPDDR $3$ (up to $1866$ MHz) \\
%$4$K video decoding at $60$ fps (H.264 or HEVC) \\
%$5$ inch, $1080\times 1920$ pixels, TFT/IPS\\
%\end{tabular} 
%\\ \hline
%\end{tabular}
%\end{center}
%\end{table}
\begin{table}[t]
\renewcommand{\arraystretch}{1.3}
\caption{Main specifications of the two tested DUTs. The displays use the in-plane switching (IPS) technology for thin-film-transistor (TFT) LCDs.  }
 \vspace{-0.4cm}
\label{tab:specs}
\begin{center}
\begin{tabular}{m{1em}|l|l}
\hline
&Module & Properties \\
\hline
%\rotatebox{90}{\textbf{Device A}} &\hspace{-0.35cm}
%\begin{tabular}{l}
%CPU \\
%\\
%\\
%GPU\\
%Memory\\
%Multimedia\\
%\end{tabular}
%& \hspace{-0.35cm}
%\begin{tabular}{l}
%Quad-core ($64$ bit) \\
% \quad $2$ low-power cores (max $1.593$ GHz)\\
%  \quad $2$ high-performance cores (max $2.15$ GHz)\\
%	Up to $624$ MHz, $319.4$ GFLOPS\\
%   $4$ GB of LPDDR $4$ (up to $1866$ MHz) \\
%$4$K video decoding at $60$ fps (H.264 or HEVC) \\
%\end{tabular} 
%\\
%\hline

% For Fairphone 2
\rotatebox{90}{\textbf{Device A}} & \hspace{-0.35cm}
\begin{tabular}{l}
CPU \\
GPU\\
Memory\\
Multimedia\\
Display\\
\end{tabular}
& \hspace{-0.35cm}
\begin{tabular}{l}
 Four $32$ bit cores (max $2.5$ GHz)\\
 Up to $578$ MHz, $147.9$ GFLOPS\\
$2$ GB of LPDDR $3$ (up to $1866$ MHz) \\
$4$K video decoding at $60$ fps (H.264 or HEVC) \\
$5$ inches, $1080\times 1920$ pixels, TFT/IPS\\
\end{tabular} 
\\
\hline
% For LG H812
\rotatebox{90}{\textbf{Device B}} &\hspace{-0.35cm}
\begin{tabular}{l}
CPU \\
\\
\\
GPU\\
Memory\\
Multimedia\\
Display\\
\end{tabular}
& \hspace{-0.35cm}
\begin{tabular}{l}
6-core\\
\quad $4$ low-power cores (max $1.4$ GHz)\\
\quad $2$ high-performance cores (max $1.8$ GHz)\\
Up to $600$ MHz, $153.6$ GFLOPS \\
 $3$ GB of LPDDR $3$ (up to $1866$ MHz) \\
 $4$K video decoding at $60$ fps (H.264 or HEVC) \\
$ 5.5$ inches, $1440\times 2560$ pixels, TFT/IPS\\
\end{tabular} 
\\
\hline
 \end{tabular}
\end{center}
\end{table}

The power model for the three tested cases \eqref{eq:PM} was trained using the sequences introduced above encoded with FFmpeg at original resolutions and crf values of $18$, $23$, $28$, and $33$, which corresponds to a `subjectively sane' range \cite{FFmpeg}. To have more information on resolutions, four sequences were encoded at quarter common intermediate format (QCIF) resolution. All in all, the set for training and validation includes $86$ single measurements. The training for optimal parameter values was performed using least-squares curve fitting. % such that the optimal parameters were determined as shown in Table\ \ref{tab:trainedValues}.
 The mean estimation error over all $M$ measurements is calculated by 
\begin{equation}
\varepsilon = \frac{1}{M}\sum_{m=1}^{M} \left| \frac{\hat P_m-P_m}{P_m}\right|,
\end{equation}
where $m$ is the measurement index, $\hat P_m$ the estimated power of the $m$-th measurement from \eqref{eq:PM}, and $P_m$ the measured power of the $m$-th measurement. To obtain the error, cross-validation was performed by separating the whole set into subsets, where each subset is based on a single input sequence. %For the separation of the training from the validation set, streams corresponding to a single input sequence were used for validation while the remaining streams were used for training.
The resulting mean errors and the trained parameter values are given in Table\ \ref{tab:trainedValues}.
 As the mean error is below $5\%$ for all tested cases, the model is highly accurate in estimating the true power of the streaming process.

%Due to the linear relation between the resolution, i.e., the number of pixels per frame, and the power (cf. \eqref{eq:PM}), highest power savings can be obtained considering HD sequences. In particular, the second largest resolution considered ($1280\times 720$) already provides less than half of the pixels per frame as HD, which also halves potential power savings related to scaling. %Therefore, in the next section, we %determine optimal resolution-crf combinations and 
 %consider power savings only for HD sequences. %More information on the training and validation process can be found in \cite{Herglotz19a}. 
\begin{table}[t]
\renewcommand{\arraystretch}{1.3}
\caption{Mean estimation error and trained parameter values for the power model \eqref{eq:PM}.  }
 \vspace{-0.4cm}
\label{tab:trainedValues}
\begin{center}
\begin{tabular}{l||c|c|c|c|c}
\hline
&$\varepsilon$ & $p_0$ & $p_f$ & $p_S$ & $p_b$ \\
\hline
A-Wi-Fi &  $3.80\%$&$1.45$ &$7.83\cdot 10^{-3}$ &$7.06\cdot 10^{-8}$ &$1.08\cdot 10^{-8}$ \\
A-3G &  $3.41\%$&$1.70$ &$7.75\cdot 10^{-3}$ &$7.40\cdot 10^{-8}$ &$1.02\cdot 10^{-8}$ \\
B-Wi-Fi &  $4.58\%$&$1.20$ &$1.30\cdot 10^{-2}$ &$5.19\cdot 10^{-8}$ &$6.22\cdot 10^{-9}$ \\
 \hline
\end{tabular}
\end{center}
\end{table}

\subsection{Results}
\begin{figure}[t]
\centering
\psfrag{014}[tc][tc]{ Estimated Power [W]}%
\psfrag{015}[bc][bc]{ PSNR [dB]}%
\psfrag{013}[bc][bc]{ }%
\psfrag{000}[ct][ct]{ $1.8$}%
\psfrag{001}[ct][ct]{ $2$}%
\psfrag{002}[ct][ct]{ $2.2$}%
\psfrag{003}[ct][ct]{ $2.4$}%
\psfrag{004}[ct][ct]{ $2.6$}%
\psfrag{005}[rc][rc]{ $24$}%
\psfrag{006}[rc][rc]{ $26$}%
\psfrag{007}[rc][rc]{ $28$}%
\psfrag{008}[rc][rc]{ $30$}%
\psfrag{009}[rc][rc]{ $32$}%
\psfrag{010}[rc][rc]{ $34$}%
\psfrag{011}[rc][rc]{ $36$}%
\psfrag{012}[rc][rc]{ $38$}%
\psfrag{data6}[bl][bl]{Pareto curve }
\psfrag{data1}[bl][bl]{$s^2 = 0.4$ }
\psfrag{data2}[bl][bl]{$s^2 = 0.6$ }
\psfrag{data3}[bl][bl]{$s^2 = 0.8$ }
\psfrag{data5}[bl][bl]{$s^2 = 0.2$ }
\psfrag{data4asdasdas}[bl][bl]{$s^2 = 1$ }
\includegraphics[width=.45\textwidth]{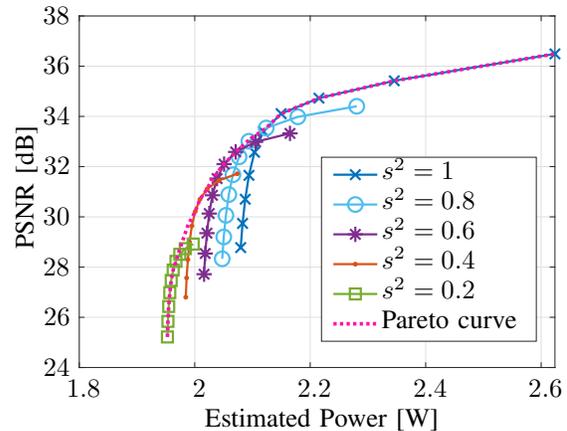}
\vspace{-.2cm}
\caption{Power-distortion curves for the BQTerrace sequence of test case A-Wi-Fi. Each curve represents a constant scaling factor as shown in the legend. Each marker on the curve represents a crf value, where the smallest value is located on the right. The Pareto curve links all Pareto efficient markers including points from the not displayed curves $s^2\in \{0.9, 0.7, 0.5, 0.3, 0.25\}$.  }
\label{fig:PD}
\end{figure}

For an exhaustive analysis, we evaluate the visual quality and the streaming power for all permutations of the downscaling factors $s^2$ and crf values $c$ shown in Table\ \ref{tab:inputVals}. 
\begin{table}[t]
\renewcommand{\arraystretch}{1.3}
\caption{Squared scaling factors $s^2$ and crf values $c$ for power-quality evaluation.  }
 \vspace{-0.4cm}
\label{tab:inputVals}
\begin{center}
\begin{tabular}{l||c|c|c|c|c|c|c|c|c|c}
\hline
$s^2$ & $1$ & ${0.9}$ & ${0.8}$ &${0.7}$ &${0.6}$ &${0.5}$ &${0.4}$ &${0.3}$ & $0.25$& $0.2$\\
\hline
$c$ & $18$ & $20$ & $22$ & $24$ & $26$ & $28$ & $30$ & $32$ & $34$ & $36$\\
 \hline
\end{tabular}
\end{center}
\end{table}
A subset of the resulting power-distortion curves depending on the coding resolution is plotted in Fig.\ \ref{fig:PD} (BQTerrace sequence, case A-Wi-Fi). 
%In the plot, we can see the distortion and the estimated streaming power for all tested combinations of scaling factors $s$ and crfs. 
Each curve corresponds to a constant scaling factor, and the markers on the curves represent crf values, where the small crf values are located on the right. 

The curves indicate that to obtain a maximum quality at low power, different combinations can be chosen. If high qualities are desired, it is helpful to choose the maximum resolution $s^2=1$ and only change the crf (blue curve). If power shall be saved, it makes sense to reduce the resolution significantly ($s^2=0.2$) and compress with medium crfs (green curve). The pink curve represents the Pareto front which links all combinations $\{s^2,c\}$ that are Pareto-efficient on a bilinearly interpolated curve. 

\begin{figure}[t]
\centering
\psfrag{013}[tc][tc]{ Estimated Power [W]}%
\psfrag{014}[bc][bc]{ VMAF Score}%
\psfrag{012}[bc][bc]{ }%
\psfrag{000}[ct][ct]{ $1.8$}%
\psfrag{001}[ct][ct]{ $2$}%
\psfrag{002}[ct][ct]{ $2.2$}%
\psfrag{003}[ct][ct]{ $2.4$}%
\psfrag{004}[ct][ct]{ $2.6$}%
\psfrag{005}[rc][rc]{ $40$}%
\psfrag{006}[rc][rc]{ $50$}%
\psfrag{007}[rc][rc]{ $60$}%
\psfrag{008}[rc][rc]{ $70$}%
\psfrag{009}[rc][rc]{ $80$}%
\psfrag{010}[rc][rc]{ $90$}%
\psfrag{011}[rc][rc]{ $100$}%
\psfrag{data6}[bl][bl]{Pareto curve }
\psfrag{data1}[bl][bl]{$s^2 = 0.4$ }
\psfrag{data2}[bl][bl]{$s^2 = 0.6$ }
\psfrag{data3}[bl][bl]{$s^2 = 0.8$ }
\psfrag{data5}[bl][bl]{$s^2 = 0.2$ }
\psfrag{data4asdasdas}[bl][bl]{$s^2 = 1$ }
\includegraphics[width=.45\textwidth]{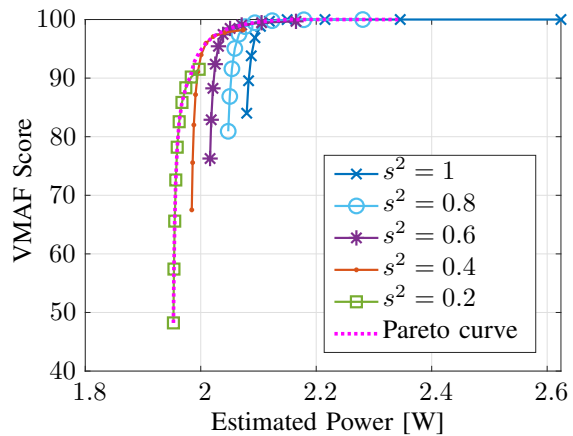}
\vspace{-.2cm}
\caption{Power-VMAF curves for the BQTerrace sequence and case A-Wi-Fi. Each curve represents a constant scaling factor as shown in the legend.  }
\label{fig:PD_VMAF}
\end{figure}

In addition to PSNR considerations, we evaluate the distortion-power curve in terms of the VMAF metric to obtain perceptual results. The curves are plotted in Fig.\ \ref{fig:PD_VMAF}. 
Again, we can see that the Pareto curve covers multiple values of the scaling factor and the crf. Interestingly, the VMAF scores for the top-right markers all yield the maximum value $100$, such that the Pareto curve does not begin with the maximum resolution and highest quality. 

\begin{figure}[t]
\centering
\psfrag{009}[tc][tc]{ Scaling Factor $s^2$}%
\psfrag{010}[bc][bc]{ Input crf $c$}%
\psfrag{000}[ct][ct]{ $0.2$}%
\psfrag{001}[ct][ct]{ $0.4$}%
\psfrag{002}[ct][ct]{ $0.6$}%
\psfrag{003}[ct][ct]{ $0.8$}%
\psfrag{004}[ct][ct]{ $1$}%
\psfrag{005}[rc][rc]{ $20$}%
\psfrag{006}[rc][rc]{ $25$}%
\psfrag{007}[rc][rc]{ $30$}%
\psfrag{008}[rc][rc]{ $35$}%
\includegraphics[width=.4\textwidth]{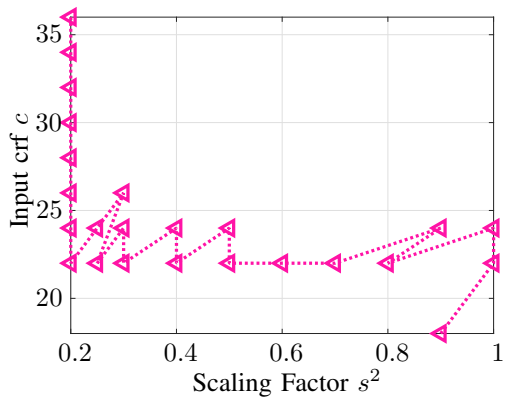}
\vspace{-.2cm}
\caption{$\{s^2,c\}$-combinations from the Pareto curve in Fig.\ \ref{fig:PD_VMAF} (BQTerrace sequence, case A-Wi-Fi). The marker at $\{s^2 = 0.9, c=18\}$ corresponds to the top right position of the Pareto curve in Fig.\ \ref{fig:PD_VMAF}.   }
\label{fig:VMAF_pareto_pairs}
\end{figure}

We take a closer look at the optimal combinations for the scaling factor and the crf $\{s^2,c\}$ by plotting the values corresponding to the Pareto curve in Fig.\ \ref{fig:VMAF_pareto_pairs}. 
The curve confirms that to obtain the highest possible VMAF quality score, it is not required to use both a maximum resolution and a minimum crf. In particular, we only observe one combination in which the crf is smaller than $22$. Furthermore, it is more helpful to reduce the resolution than using a crf larger than $24$. It is worth mentioning that with a scaling factor of $s^2=0.5$ and a crf of $22$, we still obtain a very high VMAF score of more than $98$. These results show that resolution scaling is a valid method to allow power efficient streaming at high visual qualities.

\subsection{Derivation of Recommendations}
In order to obtain a general recommendation for an optimal choice of the crf and the scaling factor, we perform a histogram analysis for all sequences introduced in Subsection\ \ref{secsec:setup} and evaluate them on all tested devices. The histogram is derived by collecting all $\{s^2,c\}$-combinations on the Pareto curve (Fig.\ \ref{fig:VMAF_pareto_pairs}). Checking the separate histograms for each test case (A-Wi-Fi, A-3G, and B-Wi-Fi), we only find minor differences such that we neglect them in the following. 

\begin{figure}[t]
\centering
\psfrag{014}[tc][tl]{ Scaling $s^2$}%
\psfrag{015}[tr][tr]{ crf values $c$}%
\psfrag{016}[bc][bc]{ Frequency of Occurrence}%
\psfrag{000}[r][r]{ $1$}%
\psfrag{001}[r][l]{ $0.8$}%
\psfrag{002}[r][l]{ $0.6$}%
\psfrag{003}[r][l]{ $0.4$}%
\psfrag{004}[r][l]{ $0.2$}%
\psfrag{005}[rc][rc]{ $18$}%
\psfrag{006}[rc][rc]{ $22$}%
\psfrag{007}[rc][rc]{ $26$}%
\psfrag{008}[rc][rc]{ $30$}%
\psfrag{009}[rc][rc]{ $34$}%%
\psfrag{010}[cr][cr]{ $0$}%
\psfrag{011}[cr][cr]{ $5$}%
\psfrag{012}[cr][cr]{ $10$}%
\psfrag{013}[cr][cr]{ $15$}%
\includegraphics[width=.49\textwidth]{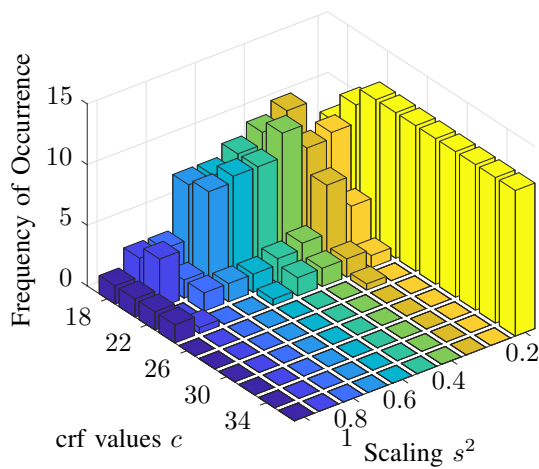}
\vspace{-.2cm}
\caption{Histogram on Pareto-efficient $\{s^2,c\}$-combinations collected for HD and $1280\times 720$-sequences from Subsection\ \ref{secsec:setup}.   }
\label{fig:hist_pairs_HD}
\end{figure}

At first, we consider HD sequences as well as sequences at the resolution $1280\times 720$ ($720$p) and plot the histogram in Fig.\ \ref{fig:hist_pairs_HD}. 
For each $\{s^2,c\}$-combination shown on the x- and y-axis, we count how often the combination occurs on the Pareto curves. For example, the combination $\{s^2=1,c=18\}$ was only Pareto efficient in three cases. In contrast, $\{s^2=0.4,c=20\}$ corresponds to a Pareto efficient combination in $21$ cases (largest green block). The yellow bars at the right correspond to very low visual qualities (the green curve in Fig.\ \ref{fig:PD} and \ref{fig:PD_VMAF}) such that we do not consider them in the following. 

The distribution of the bars in Fig.\ \ref{fig:hist_pairs_HD} suggests that to save power, downscaling at a fixed crf of $c=20$ is more beneficial than increasing $c$. As a consequence, we propose to use the combinations listed in Table\ \ref{tab:prop_comb} in practice. For these combinations, we evaluate all sequences in all test cases. As criterion, we take the mean VMAF score as well as the mean relative power savings $\Delta p$ with respect to $\{s^2=1,c=18\}$ and also list them in Table\ \ref{tab:prop_comb}. %the optimal combination  and the corresponding mean VMAF metric. 
\begin{table}[t]\renewcommand{\arraystretch}{1.3}
\caption{Recommended combinations of $\{s^2,c\}$, mean VMAF scores, and relative power savings with respect to $\{s^2=1,c=18\}$ (HD and $1280\times 720$ resolution).% The last row only shows power savings of HD sequences. 
}
 \vspace{-0.4cm}
\label{tab:prop_comb}
\begin{center}
\begin{tabular}{l||c|c|c|c|c}
\hline
$s^2$ & $1$ & ${0.9}$ & ${0.7}$ &${0.4}$  &${0.2}$\\
\hline
$c$   & $18$ & $20$     & $20$ & $20$     & $20$\\
 \hline
 VMAF & $100$ & $99.997$ & $99.88$ & $97.67$ & $89.6$ \\
$\Delta p$ ($720$p) & $0\%$ & $1.67\%$ & $2.66\%$& $4.13\%$ & $5.13\%$ \\
$\Delta p$ (HD) & $0\%$ & $7.14\%$ & $9.36\%$& $12.66\%$ & $14.81\%$ \\
\hline
\end{tabular}
\end{center}
\end{table}
We can see that with negligible losses in quality (mean VMAF score of $99.88$), power savings up to $10\%$ can be achieved for HD sequences. Accepting mean VMAF scores of around $90$, power savings up to $15\%$ can be reached. 

\begin{figure}[t]
\centering
\psfrag{013}[tc][tl]{ Scaling $s^2$}%
\psfrag{014}[tr][tr]{ crf values $c$}%
\psfrag{015}[bc][bc]{ Frequency of Occurrence}%
\psfrag{000}[r][r]{ $1$}%
\psfrag{001}[r][l]{ $0.8$}%
\psfrag{002}[r][l]{ $0.6$}%
\psfrag{003}[r][l]{ $0.4$}%
\psfrag{004}[r][l]{ $0.2$}%
\psfrag{005}[rc][rc]{ $18$}%
\psfrag{006}[rc][rc]{ $22$}%
\psfrag{007}[rc][rc]{ $26$}%
\psfrag{008}[rc][rc]{ $30$}%
\psfrag{009}[rc][rc]{ $34$}%%
\psfrag{010}[cr][cr]{ $0$}%
\psfrag{011}[cr][cr]{ $5$}%
\psfrag{012}[cr][cr]{ $10$}%
%\psfrag{013}[cr][cr]{ $15$}%
\includegraphics[width=.49\textwidth]{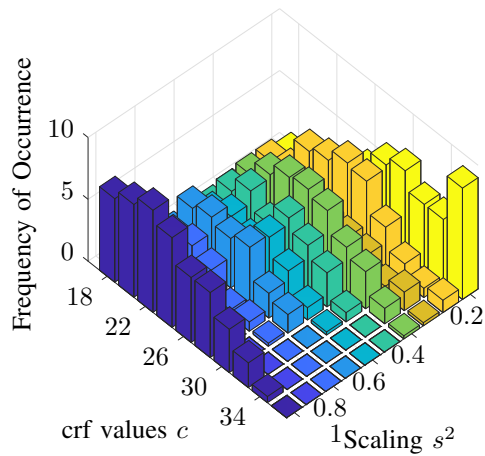}
\vspace{-.2cm}
\caption{Histogram on Pareto-efficient $\{s^2,c\}$-combinations collected for low-resolution and screen content sequences from Subsection\ \ref{secsec:setup}.   }
\label{fig:hist_pairs_LD}
\end{figure}

For the other sequences (screen content and low resolution (LR)), we plot the histogram in Fig.\ \ref{fig:hist_pairs_LD}. 
We can see that the distribution is much wider and shifted towards higher crf values. The dark blue bars at $s^2=1$ correspond to screen content, which indicates that scaling is generally not beneficial if the sequence contains high spatial frequencies (e.g., text). 

However, to determine potential power savings, we calculate the mean VMAF score and the mean relative power savings for some interesting combinations with a high frequency of occurrence, cf. Table\ \ref{tab:prop_combLR}. 
\begin{table}[t]\renewcommand{\arraystretch}{1.3}
\caption{Mean VMAF scores and relative power savings for different combinations of $\{s^2,c\}$ with respect to $\{s^2=1,c=18\}$. Values are given for low-resolution sequences (LR) and screen content. }
 \vspace{-0.4cm}
\label{tab:prop_combLR}
\begin{center}
\begin{tabular}{l||c|c|c|c|c}
\hline
$s^2$ & $1$ & ${1}$ & ${1}$ &${0.7}$  &${0.25}$\\
\hline
$c$   & $18$ & $22$     & $28$ & $22$     & $26$\\
 \hline
 VMAF (LR) & $97.53$ & $92.59$ & $77.35$ & $77.03$ & $26.24$ \\
  VMAF (screen) & $100$ & $100$ & $99.23$ & $95.75$ & $68.31$ \\
$\Delta p$ (LR)& $0\%$ & $0.81\%$ & $1.40\%$& $1.39\%$ & $2.35\%$ \\
$\Delta p$ (screen) & $0\%$ & $1.01\%$ & $1.92\%$& $2.47\%$ & $4.82\%$ \\
\hline
\end{tabular}
\end{center}
\end{table}
Apparently, maximum power savings are below $5\%$, even accepting low qualities (below a VMAF score of $70$). The reason is that due to the linear relation between the resolution, i.e., the number of pixels per frame, and the power (cf. \eqref{eq:PM}), high power savings can only be obtained by high-resolution sequences. For example, a resolution of $832\times 480$ holds less than one $5^\mathrm{th}$ of the pixels per frame as HD, which also prunes potential power savings by a factor of $5$. %Therefore, in the next section, we %determine optimal resolution-crf combinations and 
 %consider power savings only for HD sequences. %More information on the training and validation process can be found in \cite{Herglotz19a}. 
As a consequence, we conclude that spatial scaling is not beneficial for screen content as well as for resolutions of $832\times 480$ and below. 

\section{Conclusions}
\label{sec:concl}
This paper has shown that spatial scaling can be exploited to optimize the power efficiency and the perceived visual quality in mobile video streaming applications. In particular, we showed that HD streaming and videoconferencing at high resolutions can be optimized.  With the help of the proposed set of parameters, up to $10\%$ of power can be saved for near-optimal visual quality and up to $15\%$ of power can be saved for acceptable visual qualities. 

In future work, a general model for the relation between the perceived subjective quality and the resolution could be established and exploited in rate-distortion optimization. Furthermore, the sequence-specific frame rate could be considered to obtain additional power savings at decent qualities. 
%- Simulation Setup
%- Power Modeling 
%- VMAF
%- Results from simulations
%- results from measurements
%- Measurement Setup

\section*{Acknowledgment}
This work was supported by Mitacs and Summit Tech Multimedia ({https://www.summit-tech.ca/}) through the Mitacs Accelerate Program.

\bibliographystyle{IEEEtran}
\bibliography{D:/Literatur/literatureNeu}

\begin{thebibliography}{10}
\providecommand{\url}[1]{#1}
\csname url@samestyle\endcsname
\providecommand{\newblock}{\relax}
\providecommand{\bibinfo}[2]{#2}
\providecommand{\BIBentrySTDinterwordspacing}{\spaceskip=0pt\relax}
\providecommand{\BIBentryALTinterwordstretchfactor}{4}
\providecommand{\BIBentryALTinterwordspacing}{\spaceskip=\fontdimen2\font plus
\BIBentryALTinterwordstretchfactor\fontdimen3\font minus
  \fontdimen4\font\relax}
\providecommand{\BIBforeignlanguage}[2]{{%
\expandafter\ifx\csname l@#1\endcsname\relax
\typeout{** WARNING: IEEEtran.bst: No hyphenation pattern has been}%
\typeout{** loaded for the language `#1'. Using the pattern for}%
\typeout{** the default language instead.}%
\else
\language=\csname l@#1\endcsname
\fi
#2}}
\providecommand{\BIBdecl}{\relax}
\BIBdecl

\bibitem{Herglotz18a}
C.~Herglotz and A.~Kaup, ``Decoding energy estimation of an {HEVC} hardware
  decoder,'' in \emph{Proc. International Symposium on Circuits and Systems
  (ISCAS)}, Firenze, Italy, May 2018, pp. 1--5.

\bibitem{ITU_H.264}
\emph{Advanced Video Coding for Generic Audio-Visual Services}.\hskip 1em plus
  0.5em minus 0.4em\relax ITU-T Rec. H.264 and ISO/IEC 14496-10 (AVC), ITU-T
  and ISO/IEC JTC 1, {Apr.} 2003.

\bibitem{ITU_HEVC}
\emph{High Efficiency Video Coding}.\hskip 1em plus 0.5em minus 0.4em\relax
  ITU-T Rec. H.265 and ISO/IEC 23008-2, ITU-T and ISO/IEC JTC 1/SC 29/WG 11
  (MPEG), Apr. 2013.

\bibitem{statistaH264}
statista. (2019, 5) Market share of top online video codecs and containers
  worldwide from 2016 to 2018.
  https://www.statista.com/statistics/710673/worldwide-video-codecs-containers-share-online/.

\bibitem{Herglotz19a}
C.~Herglotz, S.~Coulombe, S.~Vakili, and A.~Kaup, ``Power modeling for virtual
  reality video playback applications,'' in \emph{Proc. IEEE International
  Symposium on Consumer Technology (ISCT)}, Ancona, Italy, June 2019.

\bibitem{Wang14}
R.~{Wang}, C.~{Huang}, and P.~{Chang}, ``Adaptive downsampling video coding
  with spatially scalable rate-distortion modeling,'' \emph{IEEE Transactions
  on Circuits and Systems for Video Technology}, vol.~24, no.~11, pp.
  1957--1968, Nov. 2014.

\bibitem{Afonso17}
M.~{Afonso}, F.~{Zhang}, A.~{Katsenou}, D.~{Agrafiotis}, and D.~{Bull}, ``Low
  complexity video coding based on spatial resolution adaptation,'' in
  \emph{Proc. IEEE International Conference on Image Processing (ICIP)}, Sep.
  2017, pp. 3011--3015.

\bibitem{Jenab18}
M.~{Jenab}, I.~{Amer}, B.~{Ivanovic}, M.~{Saeedi}, Y.~{Liu}, G.~{Sines}, and
  S.~{Shirani}, ``Content-adaptive resolution control to improve video coding
  efficiency,'' in \emph{Proc. IEEE International Conference on Multimedia Expo
  Workshops (ICMEW)}, July 2018, pp. 1--4.

\bibitem{Hamis19}
S.~Hamis, T.~Zaharia, and O.~Rousseau, ``Image compression at very low bitrate
  based on deep learned super-resolution,'' in \emph{Proc. IEEE International
  Symposium on Consumer Technology (ISCT)}, Ancona, Italy, June 2019.

\bibitem{Dragic14}
L.~{Dragi\'{c}}, D.~{Hofman}, M.~{Kova\v{c}}, M.~{\v{Z}agar}, and
  J.~{Knezovi\'{c}}, ``Power consumption and bandwidth savings with video
  transcoding to mobile device-specific spatial resolution,'' in \emph{Proc.
  9th International Symposium on Communication Systems, Networks Digital Sign
  (CSNDSP)}, July 2014, pp. 348--352.

\bibitem{Li12}
X.~Li, Z.~Ma, and F.~C.~A. Fernandes, ``Modeling power consumption for video
  decoding on mobile platform and its application to power-rate constrained
  streaming,'' in \emph{Proc. Visual Communications and Image Processing
  (VCIP)}, San Diego, USA, Nov. 2012.

\bibitem{Herglotz19}
C.~Herglotz, A.~Heindel, and A.~Kaup, ``Decoding-energy-rate-distortion
  optimization for video coding,'' \emph{IEEE Transactions on Circuits and
  Systems for Video Technology}, vol.~29, no.~1, pp. 172--181, Jan. 2019.

\bibitem{FFmpeg}
(2020) {Fast Forwards MPEG (FFmpeg)}. http://ffmpeg.org/. Accessed 2020-08.

\bibitem{x264}
{x264: Encoder for H.264/MPEG-4 AVC video compression}. x264.org. Accessed
  2018-11.

\bibitem{Herglotz18}
C.~Herglotz, D.~Springer, M.~Reichenbach, B.~Stabernack, and A.~Kaup,
  ``Modeling the energy consumption of the {HEVC} decoding process,''
  \emph{IEEE Transactions on Circuits and Systems for Video Technology},
  vol.~28, no.~1, pp. 217--229, Jan. 2018.

\bibitem{Ohm12}
J.-R. Ohm, G.~Sullivan, H.~Schwarz, T.~K. Tan, and T.~Wiegand, ``Comparison of
  the coding efficiency of video coding standards - including high efficiency
  video coding {(HEVC)},'' \emph{IEEE Transactions on Circuits and Systems for
  Video Technology}, vol.~22, no.~12, pp. 1669--1684, Dec. 2012.

\bibitem{Eskicioglu95}
A.~M. Eskicioglu and P.~S. Fisher, ``Image quality measures and their
  performance,'' \emph{IEEE Transactions on Communications}, vol.~43, no.~12,
  pp. 2959--2965, Dec 1995.

\bibitem{Wang04}
Z.~Wang, A.~C. Bovik, H.~R. Sheikh, and E.~P. Simoncelli, ``Image quality
  assessment: from error visibility to structural similarity,'' \emph{IEEE
  Transactions on Image Processing}, vol.~13, no.~4, pp. 600--612, Apr. 2004.

\bibitem{Zhang17}
F.~{Zhang}, A.~{Mackin}, and D.~R. {Bull}, ``A frame rate dependent video
  quality metric based on temporal wavelet decomposition and spatiotemporal
  pooling,'' in \emph{Proc. IEEE International Conference on Image Processing
  (ICIP)}, Sep. 2017, pp. 300--304.

\bibitem{VMAF}
\BIBentryALTinterwordspacing
Z.~Li, A.~Aaron, I.~Katsavounidis, A.~Moorthy, and M.~Manohara, ``Toward a
  practical perceptual video quality metric,'' \emph{The Netflix Tech Blog},
  vol.~6, 2016. [Online]. Available:
  \url{https://medium.com/netflix-techblog/toward-a-practical-perceptual-video-quality-metric-653f208b9652}
\BIBentrySTDinterwordspacing

\bibitem{Bossen13}
F.~Bossen, ``{JCTVC-L1100: Common test conditions and software reference
  configurations},'' Joint Collaborative Team on Video Coding (JCT-VC) of ITU-T
  SG16 WP3 and ISO/IEC JTC1/SC29/WG11, Geneva, Switzerland, Tech. Rep., Jan.
  2013.

\end{thebibliography}

\end{document}